\documentclass[onecolumn]{emulateapj}

\usepackage{graphicx,color,xspace}
\usepackage[fleqn]{amsmath} 
\usepackage[switch]{lineno}
\usepackage{hyperref}

\usepackage[fleqn]{amsmath}
\usepackage{amsfonts,amssymb} 
\usepackage{graphicx}
\usepackage{dcolumn, color}% Align table columns on decimal point
\usepackage{bm}% bold math
\usepackage{xparse}
\newcommand{\erf}{\ensuremath{\mathrm{erf}}}
\usepackage{natbib}
\usepackage{twoopt}
\newcommand{\GA}[1]{\ensuremath{\Gamma\left(#1\right)}}
\NewDocumentCommand\Uka{oo}{%
  \IfNoValueTF{#1}{\ensuremath{\ {_{[]}\mathcal{U}_{[]}}}}
     {\IfNoValueTF{#2}{\mbox{\ensuremath{\ _{[#1]}\mathcal{U}_{[]}}}}
       {\mbox{\ensuremath{\ _{[#1]}\mathcal{U}_{[#2]}}}}
}}
\begin{document}

\title{On the applicability of $\kappa$-distributions}
\author{K. Scherer, H. Fichtner}
\affil{Institut f\"ur Theoretische Physik IV: Plasma-Astroteilchenphysik,
  Ruhr-Universit\"at Bochum, Germany}
\affil{Research Department of Complex Plasmas,
  Ruhr-Universit\"at Bochum, Germany}
\email{kls@tp4.rub.de, hf@tp4.rub.de}
\altaffiltext{1}{Research Department, Plasmas with Complex Interactions,
  Ruhr-Universit\"at Bochum, Germany}
\author{H.~J. Fahr} 
\affil{Argelander Institut f\"{u}r Astronomie, Universit\"{a}t Bonn, Germany}
%\email{hfahr@astro.uni-bonn.de}
\author{M. Lazar} 
\affil{Centre for Mathematical Plasma Astrophysics, KU Leuven, Belgium}
\affil{Institut f\"ur Theoretische Physik IV: Plasma-Astroteilchenphysik,
  Ruhr-Universit\"at Bochum, Germany}
\begin{abstract}
  The standard (non-relativistic) $\kappa$-distribution
  is widely used to fit data and to describe macroscopic
  thermodynamical behavior, e.g.\ the pressure (temperature) as the
  second moment of the distribution function. By contrast to a
  Maxwellian distribution, for small relevant values $\kappa < 2$
  there exists  a significant, but unphysical contribution to
  the pressure from unrealistic, superluminal particles with speeds exceeding the speed of
  light. Similar concerns exist for the entropy. We demonstrate here
  that by using the recently introduced regularized
  $\kappa$-distribution one can avoid such unphysical behaviour.
\end{abstract}
\maketitle

\section{Introduction}\label{sec:1}
There are numerous data sets of particle velocity distributions exhibiting power 
law tails that can be fitted well with standard $\kappa$-distributions (SKDs, see
below), for reviews see, e.g., \citet{Pierrard-Lazar-2010}, \citet{Lazar-etal-2012} or 
\citet{Livadiotis-McComas-2013}. Limitations of the use of SKDs have been discussed 
recently by \citet{Lazar-etal-2016}, \citet{Scherer-etal-2017}, and 
\citet{Fichtner-etal-2018}. Here we discuss a further restriction to be observed when 
using SKDs. 

For various applications employing SKDs the associated temperature and pressure are needed
\citep[e.g.,][]{Heerikhuisen-etal-2008, Fahr-etal-2014, Fahr-etal-2016, Kim-etal-2018}, 
and, thus, the second-order moment of the velocity distribution. Since all moments are
calculated over the entire velocity space, which in a non-relativistic treatment extends
to infinity, this can lead to a non-negligible but unrealistic contribution from particles 
with superluminal speeds (with $v > c$, where $c = 3 \times 10^5$ km/s, is the speed of 
light in vacuum). For instance, recently Kim et al. (2018) have outlined such an unphysical 
contribution of superluminal electrons in theoretical estimations of spontaneous 
quasi-thermal emissions. Here we show that this happens in general for the pressure 
of an SKD. We further demonstrate that the regularized $\kappa$-distribution (RKD) 
introduced recently by \citet{Scherer-etal-2017} allows one to avoid this problem: A 
proper choice of the cut-off parameter in the RKD effectively truncates the superluminal
contribution, as in the case of a Maxwellian distribution. The RKD has the further 
advantage to be consistent with an extensive entropy \citep{Fichtner-etal-2018}, which 
appears not to be the case for the SKD \citep{Silva-etal-1998}.

In order to show the significance or insignificance of the contribution of superluminal 
particles to pressure and entropy we calculate `partial pressures' and 'partial 
entropies' for isotropic distribution functions defined in section~\ref{sec:2}, i.e.\
we integrate to a finite speed rather than to infinity and compare with the pressure and
entropy obtained from an integration over the entire velocity space. This is described
in section~\ref{sec:3}. After a discussion of the results in sections~\ref{sec:4} 
and~\ref{sec:5}, we discuss relevant physical systems in section~\ref{sec:6} before 
drawing conclusions in section~\ref{sec:7}.

\section{Distribution functions, pressure, and entropy}\label{sec:2}
\subsection{Distribution functions}
We discuss the following three velocity distribution functions. First, the
Maxwellian
\begin{align}
 f_{M} & = \frac{n_0}{\Theta^{3}\sqrt{\pi^{3}}}\mathrm{exp}\left({-\frac{v^{2}}{\Theta^{2}}}\right),  
\end{align}
second, the standard $\kappa$ distribution (SKD)
\begin{align}
 f_{S} &= \frac{n_0}{\Theta^{3}\sqrt{\pi^{3}\kappa^{3}}}\,\frac{\GA{\kappa+1}}{\GA{\kappa-\frac{1}{2}}}\,
          \left(1+\frac{v^{2}}{\Theta^{2}\kappa}\right)^{-\kappa-1}
\end{align}
and, third, the regularized $\kappa$ distribution (RKD) 
\begin{align}
 f_{R}(\kappa,\alpha) &=  \frac{n_0}{\Theta^{3}\sqrt{\pi^{3}\kappa^{3}}}
 \nonumber \\
 &\Uka[][0]\left(1 + \frac{v^{2}}{\Theta^{2}\kappa}\right)^{-\kappa-1}
                         \exp\left({-\alpha^{2} \frac{v^{2}}{\Theta^{2}}}\right)
\end{align}
$\Theta$
denotes the thermal speed in the Maxwellian, while for the SKD and RKD
it is a reference speed characteristic to the Maxwellian core of the
distributions \citep{Lazar-etal-2015, Lazar-etal-2016}.  For the definition of
$\Uka[][0]$
see below. The cut-off parameter $\alpha$
is process-dependent \citep[see][]{Scherer-etal-2017} and has to be
chosen accordingly. $v$
is the particle speed and $\kappa$
a parameter producing the power-law tail in the distribution function.

\subsection{Pressures}
The isotropic pressures are all calculated via the second-order moment of
the corresponding distribution functions (in spherical coordinates)
\begin{align}\label{eq:4}
  P_{i} = A \int\limits^{\infty}_{0} v^{4} f_{i}dv
\end{align}
with the $A$ containing all constant factors. The tempertaure is
defined via the ideal gas law \citep[see, e.g.\ ][]{Scherer-etal-2017,
Scherer-etal-2019}.

The corresponding pressures $P_{M}, P_{S}, P_{R}$ of the above distribution functions can 
be calculated analytically
\citep{Scherer-etal-2017,Scherer-etal-2019}:
\begin{align}
  p_{M} &= \frac{3}{2} \Theta^{2} n_{0}\\
  P_{S}(\kappa) & = \frac{3}{2} \frac{\kappa}{2\kappa-\frac{3}{2}} \Theta^{2} n_{0}\\
  p_{R}(\alpha,\kappa) &= \frac{3}{2}n_{0}\Theta^{2}\ \kappa
                             \Uka[2][0]\\
\end{align}
where $\Uka[n][m]$ is the ratio of the Kummer-$U$ (or Tricomi) functions $U(a,b,x)$:
\begin{align}
  \Uka[m][n](\kappa,\alpha) =
  \frac{U\left(\frac{3+m}{2},\frac{3+m}{2}-\kappa, \alpha^{2}\kappa\right)}
  {U\left(\frac{3+n}{2},\frac{3+n}{2}-\kappa, \alpha^{2}\kappa\right)}
\end{align}
with the empty bracket notation 
\begin{align}\nonumber
    \Uka[][n ](\kappa,\alpha) = &
   \left[{U\left(\frac{3+n}{2},\frac{3+n}{2}-\kappa, \alpha^{2}\kappa\right)}\right]^{-1}\\\nonumber
   \Uka[m][](\kappa,\alpha) %=&
 %  \frac{U\left(\frac{3+m}{2},\frac{3+m}{2}-\kappa, \alpha^{2}\kappa\right)}
 % {1}\\\nonumber
  =& U\left(\frac{3+m}{2},\frac{3+m}{2}-\kappa, \alpha^{2}\kappa\right)\\\nonumber
  \Uka[][](\kappa,\alpha) = & 1%\frac{1}{1} =                              1
\end{align}
and
\begin{align}\nonumber
   \Uka[n][n](\kappa,\alpha) = &1 
\end{align}
From the above equations \Uka[][0] and \Uka[2][0] can estimated as
functions of $\alpha$ and $\kappa$.

\subsection{Entropies}
A general definition of the entropy $S$ of a gas was given originally by
\citet{Boltzmann-1872} and \citet{Gibbs-1902} and for a plasma, e.g., by
\citet{Balescu-1975}, \citet{Balescu-1988}, and \citet{Cercignani-1988}:
\begin{eqnarray} 
\hspace*{-0.3cm}S_i = - k_B \iint f_i\, [\ln(f_i)-1]\; d^3\!r d^3\!v - k_B N \ln\left(\frac{h^3}{m^3}\right)
\label{entropy}
\end{eqnarray} 
with the phase space distribution function $f=f(\vec{r},\vec{v},t)$ of $N$ particles 
species and $h$ the Planck constant. 
This definition of the Gibbs entropy (sometimes called Boltzmann-Gibbs entropy)
is valid for both equilibrium and non-equilibrium systems, takes into account the quantum
mechanical lower limit of the phase space volume occupied by a single particle, and
 avoids the Gibbs paradoxon. For an evaluation of this 
expression for the RKD see \citet{Fichtner-etal-2018}.
\section{Partial pressures, entropies, and relative ratios}\label{sec:3}
We denote the isotropic `partial' pressures with $P'_{i}(w)$
\begin{align}
  P'_{i}(w) = A \int\limits^{w}_{0} v^{4} f_{i}dv
\end{align}
where the constant $A$, as given in Eq.~\ref{eq:4}, is not affected and $w$ is the cut-off
speed. We define the relative contribution as the ratio ($i\in\{M,S,R\}$)
\begin{align}
  R_{i} = \frac{P_{i}-P'_{i}(w)}{P_{i}} = 1- \frac{P'_{i}(w)}{P_{i}}
\end{align}
that quantifies the relative pressure contribution of particles with
speeds greater than $w$.
For a physically valid model this contribution should be negligible in
the limit $w\to c$.
In the following, we use the normalized cut-off speed $w/\Theta$,
where for physical reasons $\Theta$
is commonly associated to the thermal speed.
This immediately leads to the
condition $c/\Theta>1$ for the Maxwellian and $c/\Theta>\alpha^{-1}$ for the
RKD. For the SKD no cut-off is defined. 

In the same manner we define the partial entropies:
\begin{align}
  S'_{i} (w)&=& - 4\pi k_B \int\limits^{w}_{0}
               \left[\int f_i\, [\ln(f_i)-1]\; d^3\!r\right] v^2 dv 
%        & & - k_B N \ln\left(\frac{h^3}{m^3}\right)
\end{align}
and the relative entropy contributions
\begin{subequations}
\begin{align}
  r_{i} &= \frac{\hat{S}_{i}-S'_{i}(w)}{\hat{S}_{i}} = 1- \frac{S'_{i}(w)}{\hat{S}_{i}}
\\%\;\;\;;\;\;\;
  \hat{S}_i &= S_i + k_B N \ln\left(\frac{h^3}{m^3}\right)
\end{align}
\end{subequations}
Note, first, that only the velocity integration is 'partial' and that the integration 
w.r.t.\ position still extends over the entire configuration space. Second, we have 
omitted the constant `quantum mechanical' term involving $h$ in order to have $r_i$ as a 
direct measure of the partial contribution of the distribution function to the entropy.  
\section{The results for the pressures}\label{sec:4}
Figs.~\ref{fig:1} to \ref{fig:3} show the calculated $R_{i}$
as functions of the normalized speed $w/\Theta$. First, we discuss the
Maxwellian distribution, which is important as it is a limit for both the SKD 
and the RKD, next the SKD and finally the RKD. 
%For the latter, we also estimated the
%maxima of the partial pressures as a function of $w/\Theta$ and $\alpha$,
%which lie always in the integration range. It turns out that the maximum
%contribution to the pressure for all discussed velocity distribution functions
%comes from velocities greater than zero. The details are given below.
%
\subsection{The Maxwellian}
The results for a Maxwellian are well known but we discuss them here for a later
comparison with those for the SKD and RKD.

\begin{figure}[t!]
  \centering
  \includegraphics[width=0.5\columnwidth]{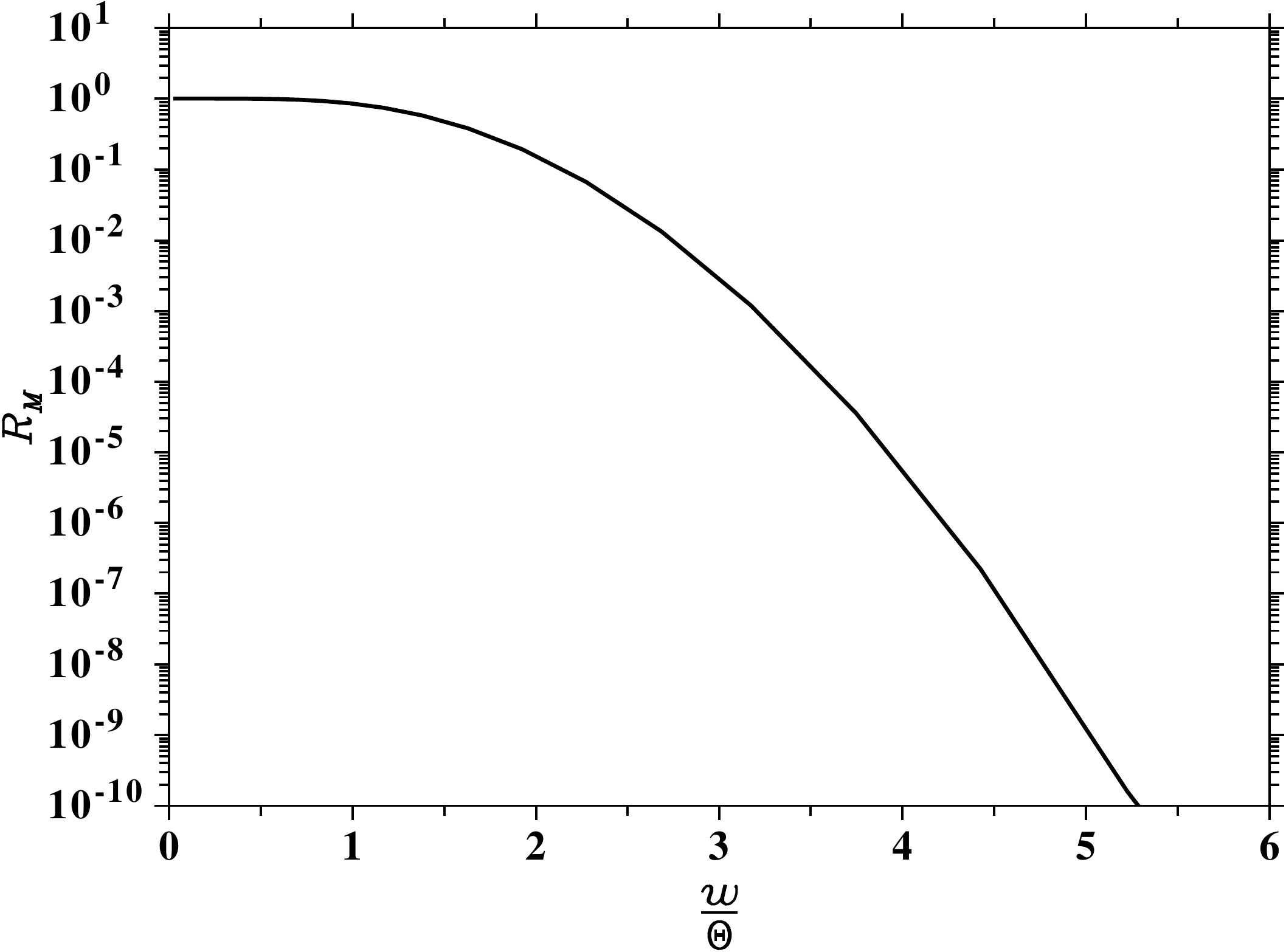}
  \caption{Plot of the relative ration $R_{M}$, i.e.\ the relative pressure contribution 
           of superluminal paricles in case of an isotropic Maxwellian velocity distribution.}
  \label{fig:1}
\end{figure}

The partial pressure for a Maxwellian is:
\begin{align}
  P'_{M}(\hat{w}) =& \frac{3}{2}\Theta^{2}n_{0}\\\nonumber
  & \left(
     \erf(\hat{w}) - \left[\frac{2\hat{w}}{\sqrt{\pi}} - \frac{4\hat{w}^{3}}{3\sqrt{\pi}}\right]\mathrm{e}^{-\hat{w}^{2}}
      \right)
\end{align}
with the above formulas one finds for the relative ratio:
\begin{align}
  R_{M} &= 1 - \erf(\hat{w})
    +\frac{1}{3\sqrt{\pi}}\left(\left[4\hat{w}^{3}+6\hat{w}\right]
             \mathrm{e}^{-\hat{w}^{2}}\right)
\end{align}
with $\hat{w}=w/\Theta$. This ratio is shown in Fig.~\ref{fig:1}. Because we have 
plotted $w/\Theta$, but not specified $\Theta$ one can read Fig.~\ref{fig:1} (and also 
Figs.~\ref{fig:2} and~\ref{fig:3}) as follows: Assuming the normalisation speed $\Theta$ is 
fixed at some value (say $\Theta=30$\,km/s for solar wind protons or $\Theta\approx 1000$\,km/s for
solar wind electrons at 1\,AU), avoiding a significant pressure contribution of particles
with superluminal speeds requires the relative ratio to become  smaller than 
$c/\Theta=10^{4}$ for protons and $c/\Theta=3\cdot10^{2}$ for electrons.

As is obvious from Fig.~\ref{fig:1}, 
$R_M$ becomes negligibly low already far below these limits. Consequently, the unphysical 
contribution of superluminal particles is negligible as well. Only if $c/\Theta$ is 
close to one, i.e.\ only if the considered particle population is characterized by 
relativistic temperatures, $R_M$ is significantly different from zero and the use of the 
non-relativistic Maxwellian is prohibited. In that case the relativistic 
Maxwell-J\"uttner distribution must be used \citep{Juettner-1911,Hakim-2011}.

\subsection{The SKD}
The partial pressure for the SKD is 
\begin{align}
 P'_{S}(\kappa,\hat{w})  =& 
    \frac{4}{5} n_{0} \Theta^{2}\,\hat{w}^{5}\frac{\GA{\kappa}}{\sqrt{k\pi}\GA{1-\frac{1}{2}}}\\\nonumber
              &    \mbox{$_{[2]}$F$_{[1]}$}\left(\left[\frac{5}{2},\kappa+1\right],
    \left[\frac{7}{2}\right],\frac{\hat{w}^{2}}{\kappa}\right)
\end{align}
where $\Gamma$ is the Gamma function and \mbox{$_{[2]}$F$_{[1]}([a,b],[c],x)$} a 
hypergeometric function. Thus, the relative ratio is:
\begin{align}
   R_{S}(\hat{w}) =& 1 - \frac{8}{15}
  \frac{\GA{\kappa-1}}{\sqrt{\pi\kappa}\GA{\kappa-\frac{3}{2}}
                   }\\\nonumber
  &\hat{w}^{5}\mbox{$_{[2]}$F$_{[1]}$}\left(\left[\frac{5}{2},\kappa+1\right],
                \left[\frac{7}{2}\right],\frac{\hat{w}^{2}}{\kappa}\right)
\end{align}

\begin{figure}[t!]
  \centering
  \includegraphics[width=0.5\columnwidth]{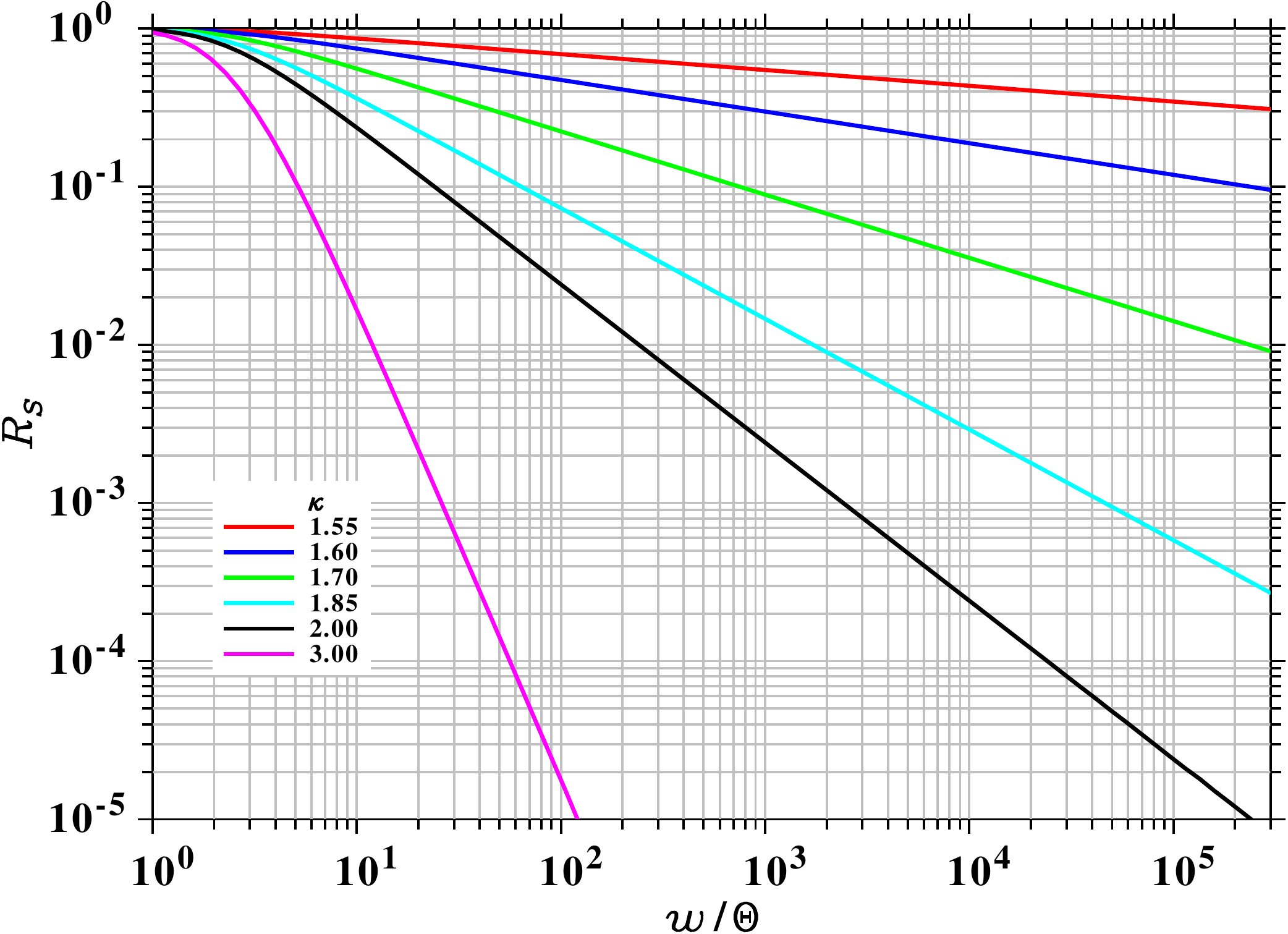}
  \caption{Plot of the relative ratio $R_{S}$ vs.\ the normalized speed $w/\Theta$ 
    for different $\kappa$ values labelled with different colors. The abcissa ends 
    at $w/\Theta=3\cdot10^{5}$, so that for $\Theta=1$\,km/s or higher reference 
    speeds one should use the gray vertical lines to determine the region of
    superluminal particle speeds, e.\ g.\ for $\Theta=1000$\,km/s all values
    $w/\Theta>3\cdot10^{2}$ lie beyond the speed of light.
  }
  \label{fig:2}
\end{figure}

In Fig.~\ref{fig:2} the relative ratio for the SKD pressure is shown for different values
of $\kappa$. The figure reveals that for, e.g., reference or thermal speed 
$\Theta=1$\,km/s the superluminal particles contribute about 1\% of the pressure for
$\kappa \leq 1.7$. This contribution becomes far more significant for higher $\Theta$,
which is more realistic for space plasmas. For example, if $\Theta = 30$\,km/s for solar
wind protons and $\Theta\approx 1000$\,km/s for electrons. As above, in the latter case
the regions of superluminal speeds are then $w/\Theta > 10^{4}$ and 
$w/\Theta > 3\cdot10^{2}$, respectively. In these cases, the contributions of 
superluminal particles to the pressure is about 3.5\% resp.\ 10\% for $\kappa=1.7$ 
and even 20\% resp.\ 40\% for $\kappa=1.6$. Given that such and even lower 
$\kappa$-values are discussed in the literature, the unphysical contribution of 
particles with speeds higher than the speed of light is often intolerably high.
While one can debate how high such a `superluminal' contribution to the pressure
can be to be tolerated, we think that one should only use the SKD for $\kappa$-values
for which $R_{S}\leq 0.01$. For lower values of $\kappa$ the relativistic SKD 
\citep{Xiao-2006} must be used. 

\subsection{The RKD}
Analogously to the previous subsections we can calculate the partial pressure and the
relative ratio for the RKD as

\begin{align}
       P'_{R}  =&
              \frac{n_{0}\Theta^{2}}{\sqrt{\pi^{3}\kappa^{3}}}\Uka[][0]
              \int\limits^{\hat{w}}_{0}\left(1 +\frac{x^{2}}{\kappa}\right)^{-\kappa-1}\mathrm{e}^{-\alpha x^{2}}x^{4}dx\\
     R_{R} =& 1 - \frac{8}{3\sqrt{\pi\kappa^{5}}}\,\Uka[][2]
  \\\nonumber
                &\int\limits_{0}^{\hat{w}}\left(1+\frac{x^{2}}{\kappa}\right)^{-\kappa-1}x^{4}\mathrm{e}^{-\alpha^{2}x^{2}} dx
\end{align}

\begin{figure*}[t!]
%  \centering
  \includegraphics[width=1.0\textwidth]{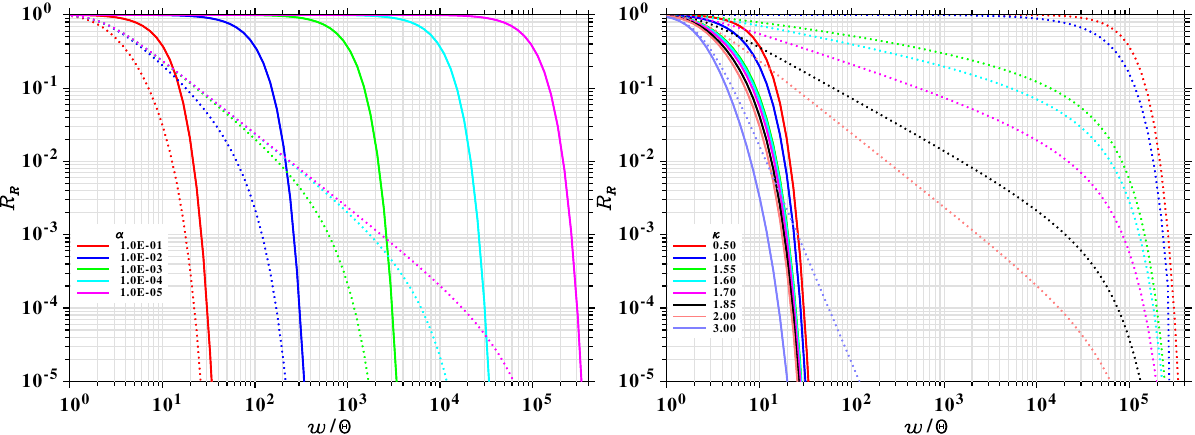}
  \caption{Left panel: The relative ratio $R_{R}$ vs.\ $w/\Theta$ for 
    $\log(\alpha) = -1,-2,-3,-4,-5$ (red, blue, green, cyan, and magenta, 
    respectively). The solid lines for $\kappa=0.5$ and the dotted lines 
    for $\kappa=2$ are the envelopes for $0.5<\kappa<2$.
    Right panel: $R_{R}$ vs.\ $w/\Theta$ for 
    $\kappa =0.5, 1, 1.55, 1.6, 1.7, 1.85, 2, 3$ (red, blue, green, cyan, magenta,
    black, light red, and light blue, respectively). The solid lines for
    $\alpha=10^{-1}$ and the dotted lines for $\alpha=10^{-5}$ are the envelops 
    for $10^{-5}<\alpha<10^{-1}$.
    }
  \label{fig:3}
\end{figure*}

In Fig.~\ref{fig:3} the contribution of the relative ratio $R_R$ is shown: While in
the left panel the parameter $\alpha$ is varied for two given $\kappa$-values
($\kappa=0.5$ and $\kappa=2$), in the right panel $\kappa$ is varied for two given values
of the cut-off parameter ($\alpha=10^{-5}$ and $\alpha=10^{-1}$). For the RKD one has the
additional requirement 
\begin{align}\label{alpha}
  \alpha > \frac{\Theta}{c}
\end{align}
in order to have an exponential cut-off at speeds lower than the speed of light $c$
\citep{Scherer-etal-2019}. While this is not a strict condition, one should choose
$\alpha$ sufficiently high so that the cut-off occurs not too close to $c$.

The left panel of Fig.~\ref{fig:3} illustrates this: for $\alpha = 10^{-5}$ and
$\kappa=0.5$ (solid magenta line) the cut-off is rather close to $c$ but the pressure 
contribution of superluminal particles is smaller than 0.01\%. The curve also reveals
that high-speed particles dominate the pressure. The higher $\alpha$ the 
smaller is this contribution, of course. For the higher value $\kappa=2$, the relative 
pressure contribution is smaller and, as expected, mainly provided by particles with 
lower speeds, i.e.\ $R_R(w/\Theta)$ has a strongly negative slope. For the above
discussed values of $\Theta=30$\,km/s and $\Theta=1000$\,km/s appropriate choices 
for $\alpha$ (see Eq.~\ref{alpha}) are $\alpha>10^{-4}$ and $\alpha>3\cdot10^{-2}$,
respectively. This is illustrated explicitly in the right panel, where for the two 
fixed values $\alpha=10^{-1}$ (solid lines) and $\alpha=10^{-5}$ (dotted lines) $\kappa$
is varied from 0.5 to 3. One can see that for $\kappa < 1$ the main contribution comes
from high-speed particles, while for $\kappa > 1$ it is provided by those with 
relatively low speeds. 
%This behavior can be understood, when one looks in the integrand for the RKD pressure:
% For $\kappa<1$ the integrand increase with a power of $w$, while for $\kappa=1$ the
% integrand is constant, and for $\kappa>1$ the integrand decreases. This behavior holds
% until the cutoff sets in. 
In cases where $\Theta$ or, correspondingly, the temperature reaches relativistic
values one has to use a relativistic version of the RKD. Such version has, to the best 
of our knowledge, not yet been formulated, in general, but is only available for the
ultra-relativistic case \citep[see][and references therein]{Treumann-Baumjohann-2018}. 
\section{The results for the entropies}\label{sec:5}
While not a moment of a distribution function, the entropy is another thermodynamic 
quantity characterizing the state of a given physical system. In order to have such 
state correctly described within a non-relativistic treatment one again can not allow
superluminal particles to have any significant contribution. As before, we quantify 
this contribution on the basis of the relative ratios defined in section~\ref{sec:3}. 

Rather than discussing all three distributions in detail again, as is done in the
previous section for the partial pressure, we illustrate the findings with the RKD
that, of course, contains the Maxwellian in the limit $\kappa\to\infty$ and that 
the SKD in the limit $\alpha\to0$. Note, however, that formula (\ref{entropy}) does
not hold for the SKD because it requires the existence of all velocity moments and,
thus, it may be necessary to use the non-extensive entropy \citep{Silva-etal-1998}.   

\begin{figure}[t!]
!  \centering
  \includegraphics[width=0.5\columnwidth]{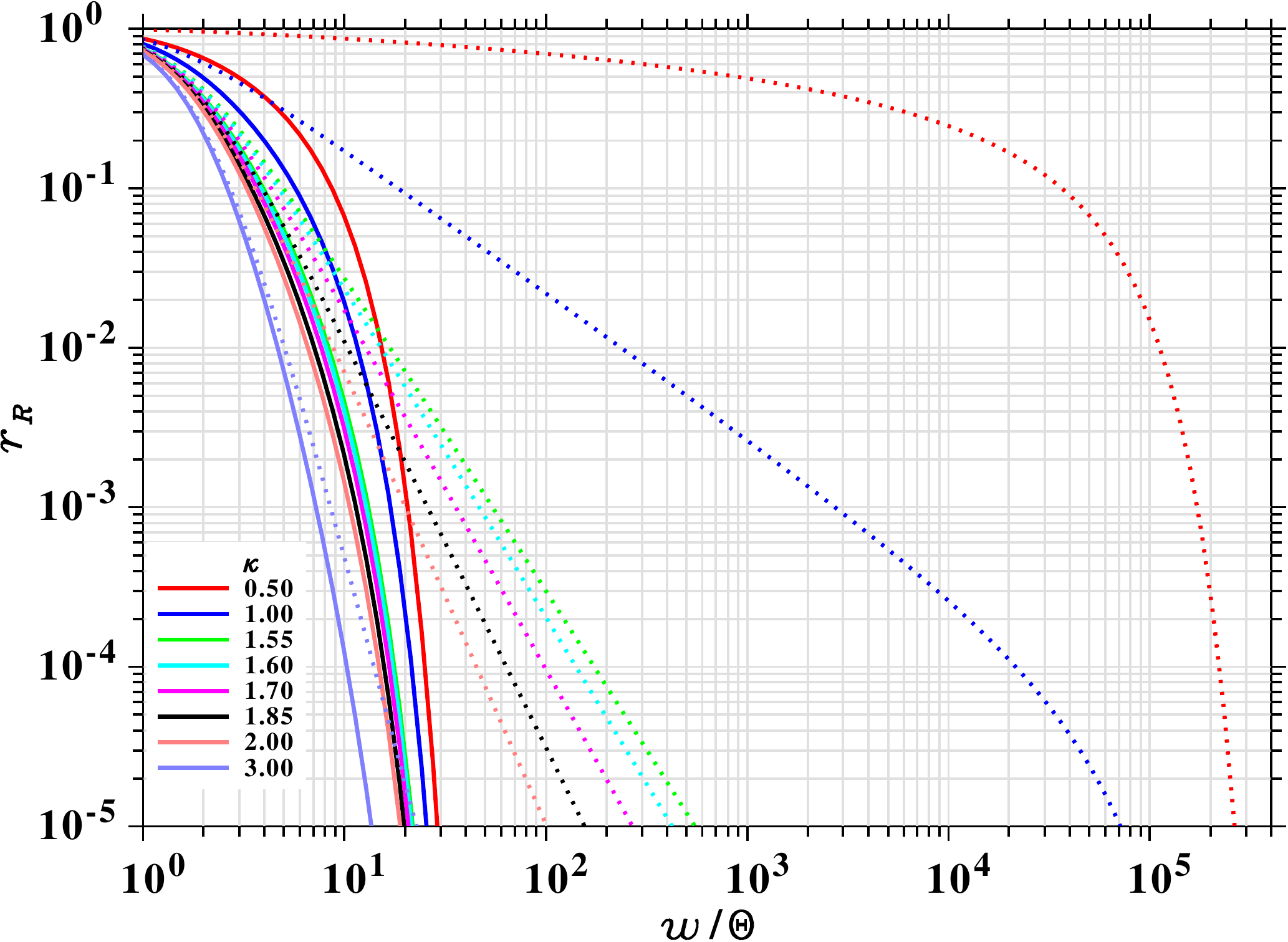}
  % \includegraphics[width=0.7\columnwidth,angle=-90]{partent-al1.pdf}
  % \centering
  % \includegraphics[width=0.7\columnwidth,angle=-90]{partent-al2.pdf}
  \caption{The relative ratio $r_{R}$ vs.\ $w/\Theta$ for 
    $\alpha = 10^{-1}$ (solid lines) and $\kappa =0.5, 1, 1.55, 1.6, 1.7, 1.85, 2, 3$
    (red, blue, green, cyan, magenta, black, light red, and light blue, respectively),
     and for $\alpha=10^{-5}$ (dotted lines).  
    }
  \label{fig:4}
\end{figure}

Fig.~\ref{fig:4} shows the relative ratio $r_{R}$ as a function
of $w/\Theta$ for all combinations of $\alpha = 10^{-5}, 10^{-1}$, and 
$\kappa =0.5, 1, 1.55, 1.6, 1.7, 1.85, 2, 3$.
The results are very similar to those for the partial pressure: First, the contributions
of particles with speeds higher than the speed of light are increasingly significant 
with decreasing $\kappa$-values. Second, chosing a higher $\alpha$-value and, thus, 
exponential cut-offs at lower speeds can reduce these contributions to insignificant
amounts. This similarity is expected, from the thermodynamic dependence of the entropy
on the pressure.  

A comparison of the results of Figs.~\ref{fig:3}~and~\ref{fig:4} 
reveals that, for given $\alpha$ and $\kappa$, the inequality $r_R < R_R$ holds, i.e.\ 
that the relative contribution of superluminal particles to the pressure is higher than
that to the entropy. Also this is expected in view of the different integrands of the 
velocity integral of the pressure moment and the entropy. 

\section{Discussion}\label{sec:6}
After we have demonstrated the possibility of the unphysical significance of 
contributions of superluminal particles to macroscopic thermodynamic quantities, 
we briefly mention three physical system for which a proper representation of the 
distribution function is essential within the framework of a non-relativistic theory.  
\subsection{Langmuir turbulence}
It was recently, shown \citep{Yoon-etal-2018} that, in contrast to the SKD, the RKD
avoids on a mesoscopic (i.e.\ kinetic) level the infrared catastrophe. The reason is
that in the long-wavelength range the contribution of the superluminal tail in case
of low $\kappa$-values is suppressed. 
\subsection{Electrons in the interplanetary medium}
The solar wind electron distribution \citep[e.g.,][]{Lin-1998} can extend to energies
beyond 100\,keV, especially in the so-called superhalo \citep{Wang-etal-2012}. For
these energies the electrons are on the edge to be in the relativistic range. 

In solar particle events there are observed power laws (in kinetic energy) with a
power-law index below 1.5 extending to even higher energies \citep{Oka-etal-2018}. 
The underlying distribution function is most likely consistent with an RKD, see 
\citet{Wang-etal-2012}. 

To avoid using the comparatively complicated relativistic $\kappa$-distribution
\citep{Xiao-2006}, one can employ for both cases the non-relativistic RKD with a 
suitable cut-off such that any contribution of superluminal electrons is neglibly small.
Moreover, cut-offs at sufficiently high velocity may prevent alterations of plasma waves dispersion
relations,  e.g., for Langmuir waves in \citet{Scherer-etal-2017} and, implicitly, unrealistic
interpretations of plasma parameters from an indirect plasma diagnosis measuring
the plasmas wave fluctuations.

\subsection{The heliosheath}
For a simulation of the heliosheath \citet{Heerikhuisen-etal-2008} and 
\citet{Fahr-etal-2016} were using the SKD with constant $\kappa=1.63$ and $\kappa < 1.7$,
respectively, which are according to the above results critical values.
While these authors do not explicitly state the value of $\Theta$, from the temperature
plot (figure~5 of the latter authors) one can estimate that $\Theta\approx 100$\,km/s in the region
beyond the termination shock. Fig.~\ref{fig:2} above reveals that for $\kappa=1.6$
and $w/\Theta<3\cdot 10^{3}$ more than 10\% of the pressure comes from superluminal
particles; for $\kappa=1.7$ this contribution reduces slightly to about 6\%.
Assuming that the effect on the heliocentric distances of the termination shock, 
the heliopause and the bow shock (given in their table~2) is of the same order, these
estimates would change accordingly. While this needs to be carefully studied again
using the RKD in the complicated charge-transfer integrals solved by
the former authors
for the SKD, our expectation is that such correction will reduce these distances. 
\section{Conclusion}\label{sec:7}
We have demonstrated that the use of non-relativistic standard $\kappa$-distributions
with low $\kappa$-values results in unphysical contributions of particles with speeds
above the speed of light to macroscopic thermodynamic quantities like pressure and
entropy. The actual limiting value of $\kappa$
depends on the thermal velocity $\Theta$
(characteristic to the Maxwellian core of such distributions) and on the
constraint adpopted for  the 'superluminal' contributions, e.\
g.\ 1\% .

While, in principle, such a limitation exists also for the non-relativistic regularized 
$\kappa$-distribution, the latter allows to suppress the unphysical significance of
superluminal particles by an appropriate choice of the cut-off parameter $\alpha$.  
This regularizing exponential cut-off makes any undesired contribution to pressure or 
entropy negligible, just as in the case of a Maxwellian distribution. Consequently, we
confirm the finding in \citet{Scherer-etal-2017} that the SKD has its merits fitting data
but can easily be inconsistent with related macroscopic quantities. Whenever there is 
the need of modelling the latter, the use of the RKD with an appropriate cut-off is 
required. 

Nevertheless, a less pragmatic and more puristic approach should aim for a consistent
formulation of relativistic versions of the SKD and the RKD. For the former this has 
been attempted by \citet{Xiao-2006}, for the latter this will be done in forthcoming 
work. 
\\
~\\

\acknowledgments
We are grateful for support from the Deutsche Forschungsgemeinschaft (DFG) via the
grants SCHE~334/9-2, SCHL~201/35-1, and from FWO-Vlaanderen (Grant GOA2316N). We also
appreciate the support from the International Space Science Institute (ISSI) for
hosting the international ISSI team on {\it Kappa Distributions: From Observational
Evidences via Controversial Predictions to a Consistent Theory of Suprathermal Space
Plasmas}, which triggered many fruitful discussions that were
beneficial for the work presented here.

%\bibliographystyle{apj}
%\bibliography{references}

\appendix
\section{Some remarks concerning the connection to other statistical
  distributions}

Although the $\kappa$
distribution (SKD) looks quite similar to the Pareto distribution used
in economics \cite[for example ][]{Arnold-2015,Krishnamoorthy-2015} or
to the Schechter luminosity ``function'' used in astronomy
\citep[e.g.\ ][]{Luo-etal-2018}, there are subtle differences: On the
first glimpse the univariate Pareto distribution is quite close to the
SKD, but because the SKD is not a univariate distribution, one has to
compare to the multivariate Pareto distribution of the fourth
kind. A statistical interpretation of the SKD or RKD is difficult,
because their (higher-order) moments have a clear physical
significance, different to those of statistical distribution functions. 

The amplitude of the velocity $|\vec{v}|=v$
depends in general on three coordinates, which reduce to one in the
case of an isotropic (spherical) distribution, because $v$
is independent on the angle variables and any integration over the solid
angle results in the factor $4\pi$.
But the moments are no longer scalars, but vectors or (higher-order) tensors
\cite[see for a more detailed
discussion][]{Scherer-etal-2019}. Because the SKD is an even function
of $v$
all odd moments vanish. Only the ``central moments'', if existing,
survive. The central moments are obtained when introducing bulk and/or
drift speed. Moreover, the probability density function already has to
be integrated over $v^{2}$
(from the volume element) which would correspond to the second-order
moment of the univariate Pareto distribution. Thus, one should be careful
comparing the Pareto distribution with the SKD or RKD.

Moreover, the SKD, and all other distribution functions in plasma
physics, must be, at least, an approximate solution of the Vlasov-,
Boltzmann-, Fokker-Planck- or Landau-equation based on the Liouville
theorem \citep{Balescu-1975,Balescu-1988,Cercignani-1988}. At least it
is known that the transport equations for cosmic ray modulation are
based on a Fokker-Planck-equation, which can be solved by stochastic
differential equations \citep[see e.g.\ ][]{Strauss-Effenberger-2017},
which are based on stochastic Wiener processes and can be extended to
Levi-flights for anomalous diffusion
\citep{Fichtner-etal-2014,Stern-etal-2014}.

%A rigorous discussion of the mathematical statistical properties would
%go far beyond the scope of this work. 

\end{document}